\documentclass[12pt,a4paper]{article}

\usepackage{authblk}
\usepackage{amsmath,amssymb}
\usepackage{graphicx,psfrag,color}
\usepackage{amsfonts}
\usepackage{epsfig}
\usepackage{times,pstricks,pst-node}
\usepackage{pstricks,pst-node,pst-text,pst-3d}
\usepackage{caption}

\newcommand{\ket}[1]{\left|#1\right\rangle}      
\newcommand{\bra}[1]{\left\langle #1\right|}     
\newcommand{\eq}{\begin{equation}}
\newcommand{\en}{\end{equation}}
\newcommand{\bear}{\begin{eqnarray}}
\newcommand{\ear}{\end{eqnarray}}

\title{Influence of boundary conditions on bulk properties of six-vertex model}
\author{T.S. Tavares$^{(1)}$\footnote{tavares@df.ufscar.br}, \ G.A.P. Ribeiro$^{(1)}$\footnote{pavan@df.ufscar.br} \ and V.E. Korepin$ ^{(2)}$\footnote{korepin@gmail.com}}
\affil{$^{(1)}$ Departamento de F\'{i}sica, Universidade Federal de S\~ao Carlos \\ S\~ao Carlos, SP 13565-905, Brazil \\
$^{(2)}$ C.N. Yang Institute for Theoretical Physics, \\
State University of New York at Stony Brook, NY 11794, USA}
\date{}

\begin{document}
\maketitle
\thispagestyle{empty}
\begin{center}
{\it In honor of Rodney Baxter's 75th birthday}
\end{center}

\begin{abstract}
We study the influence of boundary conditions on the entropy of the six-vertex model. We consider the case of fixed boundary conditions in order to argue that the entropy of the six-vertex model vary continuously from its value for ferroelectric to periodic boundary conditions. This is done by merging the ferroelectric boundary and the N\'eel boundary.
\end{abstract}


\pagestyle{plain}

\newpage

\section{Introduction}

The six-vertex model is one of the simplest and most important exactly solvable models in statistical mechanics and it has been extensively studied over the years \cite{BAXTER,BOOK}. Despite its simplicity, the six-vertex model provides a good description of the ice and spin-ice systems \cite{NATURE,SPIN-ICE}. 

This model was firstly solved with periodic boundary conditions \cite{LIEB}. Afterwards, the equivalence of the six-vertex model with free and periodic boundary conditions was shown in \cite{WU}. Additionally, it was noted that the free-energy of the six-vertex model cannot be independent of boundary conditions \cite{WU}.

The dependence of the six-vertex on boundary conditions has also been investigated. The case of special free boundaries \cite{OWCZAREK}, anti-periodic boundaries\cite{BATCHELOR} gave the same answers as the periodic boundary conditions. Later on, the six-vertex model with domain wall boundary was considered. It was proved that it produces different bulk properties in the thermodynamic limit \cite{KOREPIN2000,ZINNJUSTIN,BLEHER}, e.g the entropy at the ice-point is $S_{DW}=\frac{1}{2} \ln\left(\frac{3^3}{2^4}\right)$. Recently the case of domain wall and reflecting end boundary condition  was considered. It was also shown that the bulk properties differ from the periodic case \cite{RIBEIRO}. 

This scenario fostered a systematic investigation of the influence of boundary conditions on six-vertex model bulk properties. It was recently shown that the bulk properties depend on the boundary conditions only when one has fixed boundary \cite{TAVARES}. In other words, this implies that periodic, anti-periodic and any mixture of periodic and anti-periodic along vertical and/or horizontal direction in the rectangular lattice produce the same bulk properties. 

Nevertheless, it was also introduced in \cite{TAVARES} additional examples of fixed boundary conditions which produce different values for the entropy per lattice site. In particular, it was argued in that the entropy of the six-vertex model at the ice-point varies continuously from its value for ferroelectric boundary condition ($S_{FE}=0$) to its values with periodic boundary ($S_{PBC}=\frac{1}{2} \ln\left(\frac{4}{3}\right)^3$). However in \cite{TAVARES} it was only discussed the interval $S_{FE}< S < S_{DW}$. The purpose of this paper is to review the previous results and extend them for the whole interval $S_{FE}< S < S_{PBC}$ by means of the direct computation of the entropy per lattice site for finite lattices. This is done by merging the ferroelectric and N\'eel boundary conditions at certain fractions.

The outline of the article is as follows. In section \ref{sixvertex}, we describe the six-vertex model and its boundary conditions. In section \ref{fixed}, we discuss several instances	 of fixed boundary conditions. In section \ref{NEFEsec}, we introduce another fixed boundary condition by merging the ferroelectric and N\'eel boundary and provide some data showing that indeed the entropy is within the interval $S_{FE}<S< S_{PBC}$. Our conclusions are given in section \ref{CONCLUSION}.

\section{The six-vertex model}\label{sixvertex}

In this section, we introduce the six-vertex model and its partition function with various fixed boundary conditions.

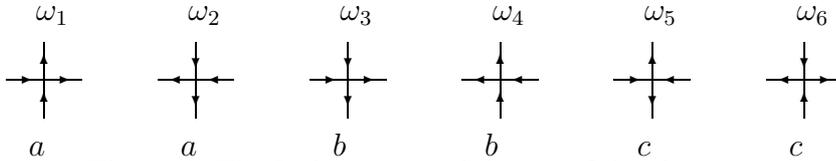
\begin{figure}[h]
\unitlength=1.0mm
\begin{center}
\begin{picture}(125,20)
\put(4,13){$\omega_1$}
\put(5,0){\line(0,1){10}}
\put(5,7.5){\vector(0,1){1}}
\put(5,2.5){\vector(0,1){1}}
\put(0,5){\line(1,0){10}}
\put(2.5,5){\vector(1,0){1}}
\put(7.5,5){\vector(1,0){1}}
\put(3,-5){$a$}
\put(24,13){$\omega_2$}
\put(25,0){\line(0,1){10}}
\put(25,7.5){\vector(0,-1){1}}
\put(25,2.5){\vector(0,-1){1}}
\put(20,5){\line(1,0){10}}
\put(22.5,5){\vector(-1,0){1}}
\put(27.5,5){\vector(-1,0){1}}
\put(23,-5){$a$}
\put(44,13){$\omega_3$}
\put(45,0){\line(0,1){10}}
\put(45,2.5){\vector(0,-1){1}}
\put(45,7.5){\vector(0,-1){1}}
\put(40,5){\line(1,0){10}}
\put(42.5,5){\vector(1,0){1}}
\put(47.5,5){\vector(1,0){1}}
\put(43,-5){$b$}
\put(64,13){$\omega_4$}
\put(65,0){\line(0,1){10}}
\put(65,2.5){\vector(0,1){1}}
\put(65,7.5){\vector(0,1){1}}
\put(60,5){\line(1,0){10}}
\put(62.5,5){\vector(-1,0){1}}
\put(67.5,5){\vector(-1,0){1}}
\put(63,-5){$b$}

\put(84,13){$\omega_5$}
\put(85,0){\line(0,1){10}}
\put(85,7.5){\vector(0,1){1}}
\put(85,2.5){\vector(0,-1){1}}
\put(80,5){\line(1,0){10}}
\put(82.5,5){\vector(1,0){1}}
\put(87.5,5){\vector(-1,0){1}}
\put(83,-5){$c$}
\put(104,13){$\omega_6$}
\put(105,0){\line(0,1){10}}
\put(105,7.5){\vector(0,-1){1}}
\put(105,2.5){\vector(0,1){1}}
\put(100,5){\line(1,0){10}}
\put(102.5,5){\vector(-1,0){1}}
\put(107.5,5){\vector(1,0){1}}
\put(103,-5){$c$}
\end{picture}
\caption{The Boltzmann weights $\omega_i$ of the six-vertex model.}
\label{6vert}
\end{center}
\end{figure}

In general, the partition function of a vertex model can be written as a sum of all configurations ($\varepsilon$),
\eq
Z=\sum_{\langle \varepsilon \rangle} \prod_{i=1}^N \prod_{j=1}^L \omega^{(i,j)}_{\varepsilon},
\label{partZ}
\en
which is a complicated combinatorial problem. The weight $\omega^{(i,j)}_{\varepsilon}$ can assume the values $a(\lambda), b(\lambda)$ and $c(\lambda)$, which are associated to the different vertex configurations of the six-vertex model (see Figure \ref{6vert})\cite{BAXTER}.

These Boltzmann weights are the matrix elements of the so called $R$-matrix \cite{BAXTER,BOOK}, which is the key ingredient for the integrability and fulfill the famous Yang-Baxter equation \cite{BAXTER}.

\section{Fixed boundary conditions}\label{fixed}

In this section we discuss some examples of boundary conditions which influence the bulk properties of the six-vertex model. We conveniently consider the case of square lattices $L=N$ throughout this work.

\subsection{Ferroelectric boundary condition}

\begin{figure}[h]
\unitlength=0.4mm
\begin{center}
\begin{picture}(100,100)(-30,-10)

\multiput(-20,0)(0,20){4}{\line(1,0){100}}
\multiput(0,-20)(20,0){4}{\line(0,1){100}}
\multiput(-8.5,0)(0,20){4}{\vector(1,0){1}}
\multiput(11.,0)(0,20){4}{\vector(1,0){1}}
\multiput(31.,0)(0,20){4}{\vector(1,0){1}}
\multiput(51.,0)(0,20){4}{\vector(1,0){1}}
\multiput(71.,0)(0,20){4}{\vector(1,0){1}}

\multiput(0,-10.)(20,0){4}{\vector(0,1){1}}
\multiput(0,11.5)(20,0){4}{\vector(0,1){1}}
\multiput(0,31.5)(20,0){4}{\vector(0,1){1}}
\multiput(0,51.5)(20,0){4}{\vector(0,1){1}}
\multiput(0,71.5)(20,0){4}{\vector(0,1){1}}

\put(-30,-1){$\lambda_4$}
\put(-30,19){$\lambda_3$}
\put(-30,39){$\lambda_2$}
\put(-30,59){$\lambda_1$}

\put(-5,85){$\mu_1$}
\put(17,85){$\mu_2$}
\put(37,85){$\mu_3$}
\put(57,85){$\mu_4$}

\end{picture}
\end{center}
\caption{The partition function $Z_N^{FE}$ for $N=4$ of the six-vertex model with ferroelectric boundary condition (FE).}
\label{fig-FE}
\end{figure}
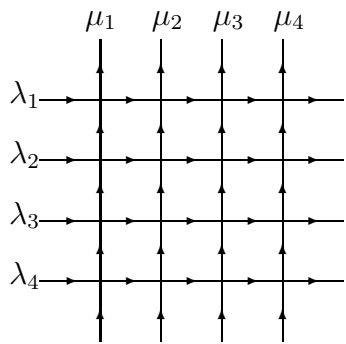

The case of ferroelectric (FE) boundary condition is trivial in the sense that one has only one allowed physical state for any finite system size (see figure \ref{fig-FE}). Therefore, the partition function is trivial for any values of the physical parameters. This is a direct consequence of the ice rule \cite{WU} and it implies that the entropy is zero ($S_{FE}=0$) .

\subsection{Domain wall boundary condition}

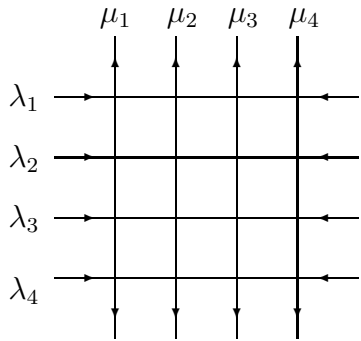
\begin{figure}[h]
\unitlength=0.4mm
\begin{center}
\begin{picture}(110,110)(-30,-10)

\multiput(-20,0)(0,20){4}{\line(1,0){100}}
\multiput(0,-20)(20,0){4}{\line(0,1){100}}
\multiput(-7.5,0)(0,20){4}{\vector(1,0){1}}
\multiput(67.5,0)(0,20){4}{\vector(-1,0){1}}
\multiput(0,-12.5)(20,0){4}{\vector(0,-1){1}}
\multiput(0,72.5)(20,0){4}{\vector(0,1){1}}

\put(-35,-7){$\lambda_4$}
\put(-35,17){$\lambda_3$}
\put(-35,37){$\lambda_2$}
\put(-35,57){$\lambda_1$}

\put(-5,85){$\mu_1$}
\put(17,85){$\mu_2$}
\put(37,85){$\mu_3$}
\put(57,85){$\mu_4$}

\end{picture}
\end{center}
\caption{The partition function $Z_N^{DW}$ for $N=4$ of the six-vertex model with domain wall boundary condition.}
\label{fig-DWBC}
\end{figure}

The first non-trivial case was introduced long ago in the context of scalar products of the Bethe states. The partition function with the domain wall (DW) boundary condition (see figure \ref{fig-DWBC}) \cite{KOREPIN1982}, 
\eq
Z_N^{DW}(\{\lambda\},\{\mu\})=\bra{\Downarrow}B(\lambda_N)\cdots B(\lambda_2) B(\lambda_1)\ket{\Uparrow},
\en
can be written as a determinant \cite{IZERGIN,KOREPIN1992}. This is a fundamental property which allowed for the computation of the bulk properties in the thermodynamic limit of the six-vertex model with DWBC\cite{KOREPIN2000,ZINNJUSTIN,BLEHER}. This also established a relation with combinatorics, which is connected with the problem of counting the number of alternating sign matrices \cite{KUPERBERG1996,ASM}.

\subsection{N\'eel boundary condition}

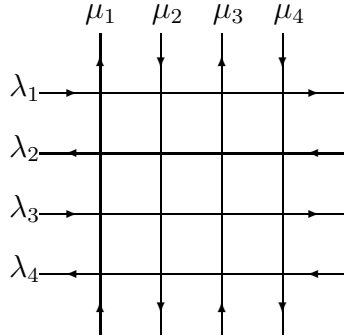
\begin{figure}[h]
\unitlength=0.4mm
\begin{center}
\begin{picture}(100,100)(-30,-10)

\multiput(-20,0)(0,20){4}{\line(1,0){100}}
\multiput(0,-20)(20,0){4}{\line(0,1){100}}
\multiput(-10.5,0)(0,40){2}{\vector(-1,0){1}}
\multiput(-8.5,20)(0,40){2}{\vector(1,0){1}}
\multiput(69.,0)(0,40){2}{\vector(-1,0){1}}
\multiput(71.,20)(0,40){2}{\vector(1,0){1}}

\multiput(20,-12.)(40,0){2}{\vector(0,-1){1}}
\multiput(0,-10.)(40,0){2}{\vector(0,1){1}}

\multiput(20,69.5)(40,0){2}{\vector(0,-1){1}}
\multiput(0,71.5)(40,0){2}{\vector(0,1){1}}

\put(-30,-1){$\lambda_4$}
\put(-30,19){$\lambda_3$}
\put(-30,39){$\lambda_2$}
\put(-30,59){$\lambda_1$}

\put(-5,85){$\mu_1$}
\put(17,85){$\mu_2$}
\put(37,85){$\mu_3$}
\put(57,85){$\mu_4$}

\end{picture}
\end{center}
\caption{The partition function $Z_N^{NE}$ for $N=4$ of the six-vertex model with N\'eel boundary condition (NE).}
\label{fig-NE}
\end{figure}

Recently, it was introduced the case of called N\'eel boundary condition or anti-ferroelectric boundary \cite{TAVARES}. This is the case where we have the alternation of the arrows along the boundaries, see Figure \ref{fig-NE}.

In contrast with the case of ferroelectric boundary condition which allows for only one possible state, the N\'eel boundary is the one which allows for the largest number of configurations. As a consequence of the arrows alternation in the boundary, it allows the largest number of arrow reversals along the boundary and this propagates to the bulk. It was shown in \cite{TAVARES} that $S_{NE}=S_{PBC}$ at ice-point ($a=b=c=1$) in the thermodynamic limit.

However, it was not possible to derive a product formula for the number of states involving factorials for the N\'eel boundary. The only estimates was obtained from the data for finite system size up to $N=20$, which indicates that the entropy behaves as $S_{NE}=S_{PBC} (1 -\frac{\gamma}{N})$, where $\gamma \sim 2$ \cite{TAVARES}. Interesting enough, the N\'eel boundary conditions also appears in the context of generalized alternating sign matrices\cite{BRUALDI}.

\subsection{Merge of DW and FE boundary condition}

We can generate additional boundaries by merging the previous cases. This results in different values for the bulk properties. By merging the domain wall and ferroelectric boundary, we have a smaller number of physical states and therefore the entropy is smaller than the domain wall boundary.
\begin{figure}[ht]
\unitlength=0.4mm
\begin{center}
\begin{picture}(130,130)(-30,-10)

\multiput(-20,0)(0,20){5}{\line(1,0){120}}
\multiput(0,-20)(20,0){5}{\line(0,1){120}}
\multiput(-7.5,40)(0,20){3}{\vector(1,0){1}}
\multiput(89,0)(0,20){5}{\vector(-1,0){1}}
\multiput(0,-12.5)(20,0){3}{\vector(0,-1){1}}
\multiput(0,92.5)(20,0){5}{\vector(0,1){1}}

\put(-40,82){\vector(0,-1){44}}
\put(-46,58){$n$}
\put(-2,112){\vector(1,0){44}}
\put(15,115){$n$}

\multiput(-9,0)(0,20){2}{\vector(-1,0){1}}
\multiput(60,-12)(20,0){2}{\vector(0,1){1}}

\multiput(69,0)(0,20){5}{\vector(-1,0){1}}
\multiput(49,0)(0,20){5}{\vector(-1,0){1}}
\multiput(29,0)(0,20){2}{\vector(-1,0){1}}
\multiput(9,0)(0,20){2}{\vector(-1,0){1}}

\multiput(0,9)(20,0){3}{\vector(0,-1){1}}
\multiput(0,29)(20,0){3}{\vector(0,-1){1}}

\multiput(60,11)(20,0){2}{\vector(0,1){1}}
\multiput(60,31)(20,0){2}{\vector(0,1){1}}
\multiput(60,51)(20,0){2}{\vector(0,1){1}}
\multiput(60,71)(20,0){2}{\vector(0,1){1}}

\put(-35,-3){$\lambda_5$}
\put(-35,17){$\lambda_4$}
\put(-35,37){$\lambda_3$}
\put(-35,57){$\lambda_2$}
\put(-35,77){$\lambda_1$}

\put(-5,105){$\mu_1$}
\put(17,105){$\mu_2$}
\put(37,105){$\mu_3$}
\put(57,105){$\mu_4$}
\put(77,105){$\mu_5$}

\end{picture}
\end{center}
\caption{The partition function $Z_N^{DW-FE}$ for $N=5$ of the six-vertex model whose boundary are a mixture of the domain wall and ferroelectric boundary (DW-FE).}
\label{fig-fDWBC}
\end{figure}
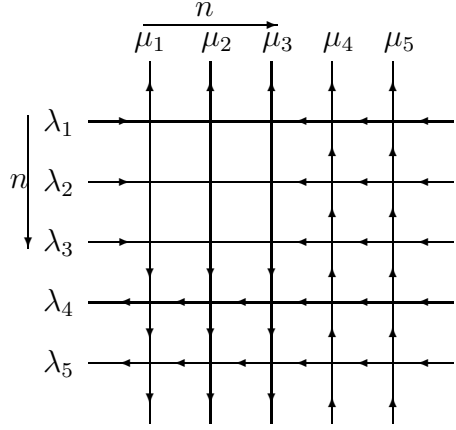

We build that by choosing an integer number $n$ between $0$ and $N$. Starting from the upper-left corner, we fill the first $n$ boundary row and column edges with arrows in the same way we would fill the domain wall boundary, see Figure \ref{fig-fDWBC}. The opposite edges of these are also filled with the respective arrows of the opposite edges of the domain wall boundary condition. So far, we have used the arrows configuration of a domain wall boundary with lattice size $n$ to fill our boundary of lattice size $N$. The remaining arrows are filled in the same way, but using boundary arrows of the ferroelectric boundary.

This implies that we have partially frozen the arrow configurations of the lattice in a similar way as the ferroelectric boundary. The difference lies in the $n\times n$ sublattice at the upper-left corner. This implies we are left with a domain wall partition function of size $n$, which means 
\eq
Z_N^{DW-FE}= \left[\prod_{i=1}^{N}\prod_{j=n+1}^{N} b(\lambda_i-\mu_j)\right]\left[ \prod_{i=n+1}^{N}\prod_{j=1}^{n} a(\lambda_i-\mu_j)\right]\times Z_n^{DW}.
\en
Therefore, we see that the entropy at infinity temperature (ice-point) is given by
\eq
S_{DW-FE}=\lim_{N \rightarrow \infty }{\left(\frac{n}{N} \right)}^2 S_{DW}.
\en
For a suitably chosen sequence $n(N)$, one can obtain any value of entropy $S$ in the interval $S_{FE} \leq S \leq S_{DW}$ \cite{TAVARES}. Therefore, the merge of domain wall and ferroelectric boundary condition implies that the entropy vary from its values for the ferroelectric case to the domain wall boundary case. However this leaves the interval $S_{DW} \leq S \leq S_{PBC}$ as an open problem.

\section{Merge of N\'{e}el and FE boundary}\label{NEFEsec}

\begin{figure}[ht]
\unitlength=0.4mm
\begin{center}
\begin{picture}(130,150)(-30,-10)

\multiput(-20,0)(0,20){6}{\line(1,0){140}}
\multiput(0,-20)(20,0){6}{\line(0,1){140}}

\put(-40,102){\vector(0,-1){64}}
\put(-46,68){$n$}
\put(-2,135){\vector(1,0){64}}
\put(15,137){$n$}

\put(-7.5,100){\vector(1,0){1}}
\put(-9.5,80){\vector(-1,0){1}}
\put(-7.5,60){\vector(1,0){1}}
\put(-9.5,40){\vector(-1,0){1}}
\put(-7.5,20){\vector(1,0){1}}
\put(-7.5,0){\vector(1,0){1}}

\put(110,100){\vector(1,0){1}}
\put(109,80){\vector(-1,0){1}}
\put(110,60){\vector(1,0){1}}
\put(109,40){\vector(-1,0){1}}
\put(110,20){\vector(1,0){1}}
\put(110,0){\vector(1,0){1}}

\put(0,-12.5){\vector(0,1){1}}
\put(20,-14.5){\vector(0,-1){1}}
\put(40,-12.5){\vector(0,1){1}}
\put(60,-14.5){\vector(0,-1){1}}
\put(80,-14.5){\vector(0,-1){1}}
\put(100,-14.5){\vector(0,-1){1}}

\put(0,110){\vector(0,1){1}}
\put(20,108){\vector(0,-1){1}}
\put(40,110){\vector(0,1){1}}
\put(60,108){\vector(0,-1){1}}
\put(80,108){\vector(0,-1){1}}
\put(100,108){\vector(0,-1){1}}

\put(80,9){\vector(0,-1){1}}
\put(100,9){\vector(0,-1){1}}

\put(80,27.5){\vector(0,-1){1}}
\put(100,27.5){\vector(0,-1){1}}

\put(72,0){\vector(1,0){1}}
\put(92,0){\vector(1,0){1}}
\put(72,20){\vector(1,0){1}}
\put(92,20){\vector(1,0){1}}

\put(-35,-3){$\lambda_6$}
\put(-35,17){$\lambda_5$}
\put(-35,37){$\lambda_4$}
\put(-35,57){$\lambda_3$}
\put(-35,77){$\lambda_2$}
\put(-35,97){$\lambda_1$}

\put(-5,125){$\mu_1$}
\put(17,125){$\mu_2$}
\put(37,125){$\mu_3$}
\put(57,125){$\mu_4$}
\put(77,125){$\mu_5$}
\put(97,125){$\mu_6$}
\end{picture}
\end{center}
\caption{The partition function $Z_N^{NE-FE}$ for $N=6$ and $n=4$ of the six-vertex model which is a merge of the N\'eel and ferroelectric boundary (NE-FE).}
\label{figNEFEfusion}
\end{figure}
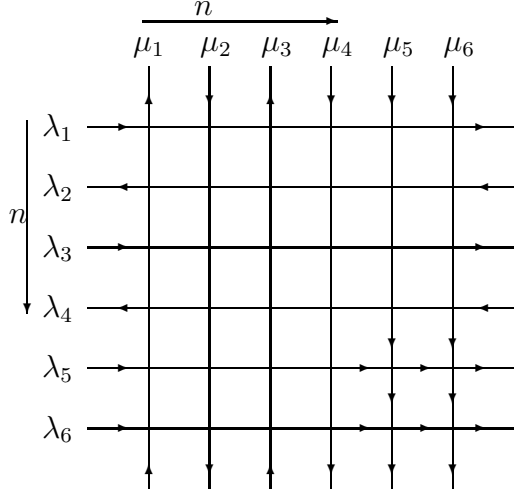

We consider the merge of the N\'eel and ferroelectric boundary in order to show that the entropy of the six-vertex model can vary in the whole interval $S_{FE} \leq S \leq S_{PBC}$. We have chosen the upper-left corner to be of N\'{e}el type and the lower-right corner to be of ferroelectric type, Figure \ref{figNEFEfusion}.

Taking into account the internal frozen degrees of freedom, one finds that the partition function takes the form of a L-shaped lattice rather than the usual rectangular one. This type of partition function have been investigated in the literature\cite{COLOMO} with different boundary conditions. 

\begin{figure}[t!]
\begin{center}
\includegraphics[width=0.8\linewidth]{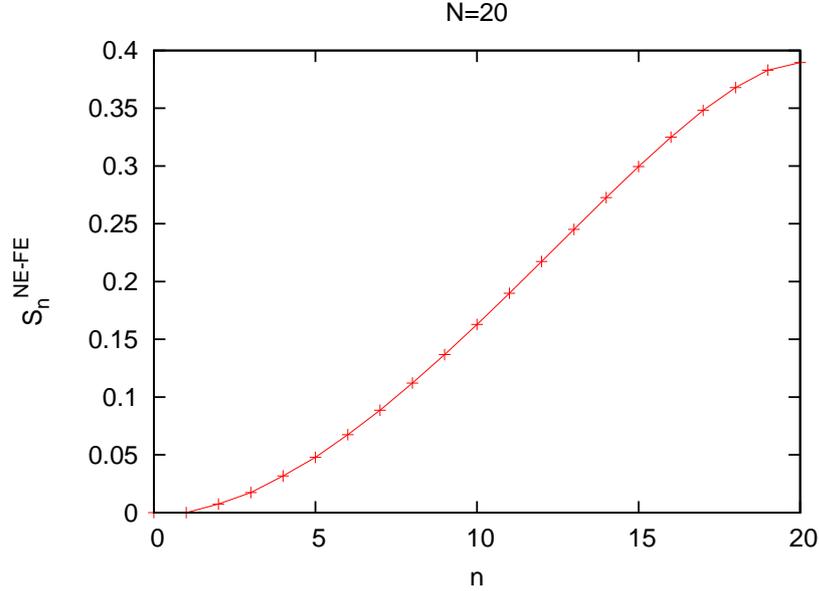}
\end{center}
\caption{Finite entropy of N\'{e}{e}l-Ferroelectric merge vs. proportion number $n$. The linear length is set to $N =20$. Note the gradual variation on the entropy values.}
\label{fig22}
\end{figure}

By construction, we have that $S^{NE-FE}=S^{FE}=0$ for $n=0$ and $S^{NE-FE}=S^{NE}=S^{PBC}$ for $n=N$ which holds true in the thermodynamic limit. We argue that if $n$ changes from $n_0$ to $n_0+1$ there is a small variation in the entropy, as expected for large $N$ values. This is supported in Figure \ref{fig22}, where we show the entropy with fixed $N=20$ and $n$ ranging from $0$ to $N$.

\begin{figure}
\begin{center}
\includegraphics[width=0.7\columnwidth]{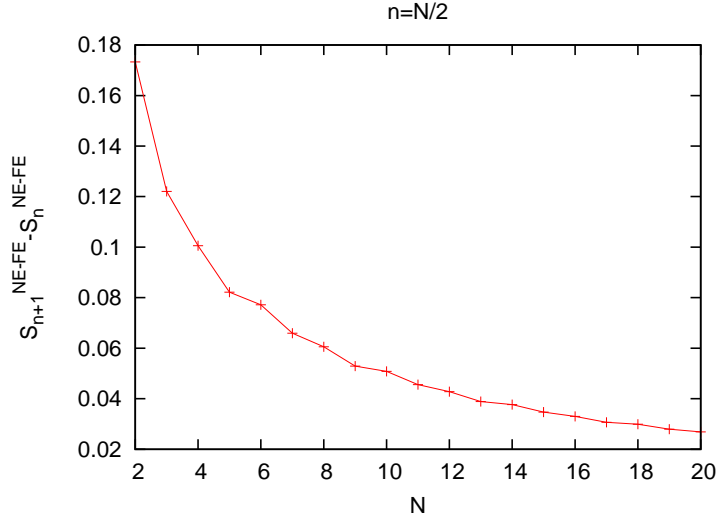}
\end{center}
\caption{Difference of finite entropies of N\'{e}{e}l-Ferroelectric merge with consecutive $n$ values vs. total linear length $N$. We set the proportion number $n_<$ to $\lfloor\frac{N}{2}\rfloor$ and $n_>=n_<+1$.}
\label{fig23}
\end{figure}

With $N$ as large as $20$, we already can see the continuity taking place. The largest difference of consecutive entropies is less than $0.03$. It reasonable to assume that $S_{n+1}^{NE-FE}-S_{n}^{NE-FE}= O(\frac{1}{N})$. To support this, we also show the entropy difference $S_{n+1}^{NE-FE}-S_{n}^{NE-FE}$ with $n=\lfloor \frac{N}{2}\rfloor$ and $N$ ranging from $2$ to $20$, Figure \ref{fig23}. As we can see, the entropy difference vanishes sufficiently fast.

\section{Open problems}

\begin{itemize}
 \item Classify the  boundaries conditions into classes, which produce the same entropy per lattice site in thermodynamic limit.
 \item Prove that the bulk free energy is  constant  inside of each class.
 \item Find new boundary conditions, for which the model is solvable analytically.
 \item Prove that for majority  of boundary evaluation of bulk free energy is NP hard.
 \item Prove that for majority of boundary conditions the phase boundaries [in the space of Boltzmann weights] are the same as for periodic case.
\end{itemize}

\section{Conclusion}
\label{CONCLUSION}

In this paper we have studied the dependence of physical quantities, like entropy, of the six-vertex model on boundary conditions.

We argued that the entropy per lattice site changes continuously from zero to its value with periodic boundary condition in the thermodynamic limit. 

There still remains open questions, e.g the complete classification of the boundary conditions, the existence of further boundary conditions for which the model is solvable analytically and the existence of other vertex models whose physical quantities do depend on the boundary conditions.

\section*{Acknowledgments}
T.S. Tavares and G.A.P. Ribeiro thank the Galileo Galilei Institute and the organizers of the scientific program "Statistical Mechanics, Integrability and Combinatorics" for hospitality and support during part of this work and the S\~ao Paulo Research Foundation (FAPESP) for financial support through the grants 2013/17338-4 and 2015/01643-8. V.E. Korepin was supported by NSF Grant DMS 1205422.

\end{document}